\documentclass[pra,reprint,a4paper,showpacs]{revtex4-1}
\usepackage{bm}
\usepackage{ulem}
\usepackage[dvipdfmx]{graphicx}
\usepackage{dcolumn}
\usepackage{color}
\usepackage{amsmath}
\usepackage{amssymb}
\usepackage{amsfonts}

\newcommand{\ue}[1]{$^{\textrm{#1}}$}

\draft 

\begin{document}


\title{Gauge-invariant formulation of time-dependent configuration interaction singles method
}



\author{Takeshi Sato\ue{1,2}}
\email[Electronic mail:]{sato@atto.t.u-tokyo.ac.jp}
\author{Takuma Teramura\ue{1}}
\email[Electronic mail:]{teramura@atto.t.u-tokyo.ac.jp}
\author{Kenichi L. Ishikawa\ue{1,2}}
\email[Electronic mail:]{ishiken@n.t.u-tokyo.ac.jp}
\affiliation{
\ue{1}Department of Nuclear Engineering and Management, School of Engineering,
The University of Tokyo, 7-3-1 Hongo, Bunkyo-ku, Tokyo 113-8656, Japan
}
\affiliation{
\ue{2}Photon Science Center, School of Engineering, 
The University of Tokyo, 7-3-1 Hongo, Bunkyo-ku, Tokyo 113-8656, Japan
}




\begin{abstract}
We propose a gauge-invariant formulation of the channel
orbital-based time-dependent configuration interaction singles (TDCIS)
method [Phys.~Rev.~A, {\bf 74}, 043420 (2006)], one of the powerful {\it
ab initio} methods to investigate electron dynamics in atoms and
molecules subject to an external laser field.
In the present formulation, we derive the equations of motion
(EOMs) in the velocity gauge using gauge-transformed orbitals, {\it not
fixed orbitals}, that are equivalent to the conventional EOMs in the
length gauge using fixed orbitals. The new velocity-gauge EOMs
avoid the use of the length-gauge dipole operator, which diverges at
large distance,
and allows to exploit computational advantages of the
velocity-gauge treatment over the length-gauge one, e.g, a faster
convergence in simulations with intense and long-wavelength lasers, and
the feasibility of exterior complex scaling as an absorbing boundary.
The reformulated TDCIS method is applied to an exactly solvable model
of one-dimensional helium atom in an intense laser field to
numerically demonstrate the gauge invariance.
We also discuss the consistent method for evaluating the time derivative of
an observable, relevant e.g, in simulating high-harmonic generation.
\end{abstract}


\maketitle 


\section{Introduction\label{sec:introduction}}
Time-dependent configuration interaction singles (TDCIS) method is one of the powerful
 {\it ab initio} methods to investigate laser-driven electron dynamics in
 atoms and molecule \cite{Gordon:2006,Rohringer:2006,Rohringer:2009,Greenman:2010,Pabst:2011,Pabst:2012,Sytcheva:2012,Pabst:2012b,Pabst:2013,Pabst:2013b,Heinrich-Josties:2015,Pabst:2014,Hollstein:2015,You:2016,You:2017,Grosser:2017,Krebs:2014,Karamatskou:2014,Chen:2015,Tilley:2015,Karamatskou:2015,Goetz:2016,Karamatskou:2017,Karamatskou:2017b}.
In the TDCIS method, the time-dependent electronic wavefunction is given
by the configuration interaction (CI) expansion, 
\begin{eqnarray}\label{eq:tdcis_1q}
\Psi(t) = \Phi C_0(t) + \sum_i^{occ}\sum_a^{vir} \Phi_{ia} C_{ia}(t),
\end{eqnarray}
where $\Phi$ is the ground-state Hartree-Fock (HF) wavefunction, and
$\Phi_{ia}$ is a singly-excited configuration-state function (CSF), replacing
an occupied HF orbital $\phi_i$ in $\Phi$ with a virtual 
(unoccupied in $\Phi$) orbital $\phi_a$, and
the electron dynamics is described through the time evolution of the CI
coefficients, $C_0$ and $\{C_{ia}\}$.
Compared to more involved {\it ab initio} wavefunction-based approaches
\cite{Ishikawa:2015} such as time-dependent multiconfiguration
self-consistent-field (TD-MCSCF) methods
\cite{Zanghellini:2003,Kato:2004,Caillat:2005,Miyagi:2013,Sato:2013,Miyagi:2014b,Haxton:2015,Sato:2015}, 
time-dependent $R$-matrix based approaches \cite{Burke1997JPB,Lysaght2008PRL,Lysaght2009PRA}, or time-dependent
reduced density-matrix approach\cite{Lackner:2015,Lackner:2017}, 
distinct advantages of the TDCIS method include a low computational cost and the conceptual simplicity to analyze simulation results. 
Furthermore, an equivalent, effective one-electron theory with coupled
channels has been developed \cite{Rohringer:2006}, which introduces the orbital-like
quantity, called channel orbital, 
\begin{eqnarray}\label{eq:chi_1q}
\chi_i(\bm{r},t) = \sum_a \phi_a(\bm{r}) C_{ia}(t),
\end{eqnarray}
and rewrites EOMs for CI coefficients with those for channel orbitals
$\{\chi_i(\bm{r},t)\}$ with no reference to virtual orbitals. This reformulation removes the
bottleneck of the CI coefficient-based TDCIS method to compute all (or, at least sufficiently many, including bound and continuum) virtual
orbitals prior to the simulation, and thus particularly useful in grid-based simulations.

Despite this advantage, numerical applications of the channel
orbital-based TDCIS method has been limited to
Refs.~\cite{Rohringer:2006,You:2016,You:2017} for a one-dimensional Hamiltonian and Ref.~\cite{Gordon:2006}
for noble gas atoms with a Hartree-Slater potential, as far as we know, and the vast
majority of applications to date have adopted the CI
coefficient-based approach \cite{Rohringer:2009,Greenman:2010,Pabst:2011,Pabst:2012,Sytcheva:2012,Pabst:2012b,Pabst:2013,Pabst:2013b,Heinrich-Josties:2015,Pabst:2014,Hollstein:2015,You:2016,You:2017,Grosser:2017,Krebs:2014,Karamatskou:2014,Chen:2015,Tilley:2015,Karamatskou:2015,Goetz:2016,Karamatskou:2017,Karamatskou:2017b}
, except for the use of $\{\chi_i\}$ as intermediate quantities in evaluating photoelectron spectra \cite{Karamatskou:2014}.
The preference of CI coefficient-based approach might be partially due
to the high symmetry of atomic systems, for which the stationaly
Hartree-Fock operator decouples for different angular momenta
\cite{Greenman:2010}, making it a relatively feasible task to obtain 
all virtual orbitals (within a given radial grids or radial basis functions) for 
{\color{black}the} lowest few angular momenta. The channel orbital-based approach would be more suited, on the
other hand, to simulations of electron dynamics with intense and/or long-wavelength laser
fields, requiring much longer angular momentum expansion \cite{Nurhuda:1999,Grum-Grzhimailo:2010,Sato:2016}, and moreover
to grid-based molecular applications, where obtaining a sufficient spectrum of virtual levels could be unacceptably expensive. 

However, the TDCIS method, either in the CI coefficient-based or channel
orbital-based formulation, suffers from the lack of gauge invariance, as a general consequence of
relying on truncated CI expansion with fixed orbital functions.
Previously, the length gauge (LG) has been employed e.g, in
Ref.~\cite{Rohringer:2006,Rohringer:2009,Greenman:2010,Pabst:2011,Pabst:2012,Sytcheva:2012,Pabst:2012b,Pabst:2013,Pabst:2013b,Heinrich-Josties:2015,Pabst:2014,Hollstein:2015,You:2016,You:2017,Grosser:2017}, and the velocity gauge (VG) in
Ref.~\cite{Krebs:2014,Karamatskou:2014,Chen:2015,Tilley:2015,Karamatskou:2015,Goetz:2016,Karamatskou:2017,Karamatskou:2017b}.
Although gauge dependence of the TDCIS method using fixed orbitals has been noted already in
Ref.~\cite{Rohringer:2006}, comparative assessment of the LG and
VG treatments (within the grid-based TDCIS) has not been reported to the best of our knowledge,
except {\color{black}for} being briefly mentioned in Ref.~\cite{Dahlstrom:2017}.
In particular, the channel orbital-based approach \cite{Rohringer:2006}
has been applied only in the LG \cite{Gordon:2006,Rohringer:2006}, and
as shown below in this paper, the VG treatment with fixed orbitals is
not very appropriate for applications to high-field phenomena. This is a
serious drawback, since for an efficient simulation of molecules, it is
highly appreciated to take advantage of the velocity-gauge treatment, 
e.g, the feasibility of exterior complex scaling
\cite{Scrinzi:2010,Orimo:2018} as an absorbing boundary, 
to reduce the computational cost related to the number of grid points.

In the present work, we propose a gauge-invariant reformulation of the
channel orbital-based TDCIS method. To this end, instead of applying the
fixed-orbital TDCIS ansatz to the velocity-gauge time-dependent
Schr{\"{o}}dinger equation (TDSE), we adopt the formulation 
using unitary-rotated orbital 
$\phi^\prime_p(t)=U(t)\phi_p$, where $U(t)$ is the gauge transformation
operator connecting the (exact) solution of TDSE in the LG and VG.
The resulting EOMs in the reformulated VG is equivalent to the LG ones with fixed orbitals by
construction, and at the same time allows to exploit advantages of the
velocity-gauge simulations as mentioned above. 

This paper proceeds as follows. In Sec.~\ref{sec:theory}, after
defining the target Hamiltonian and the gauge transformation in
Sec.~\ref{subsec:system} and reviewing the TDCIS method using fixed
orbitals both in the CI coefficient-based
[Sec.~\ref{subsec:tdcis_fixed_cicoeff}] and channel orbital-based [Sec.~\ref{subsec:tdcis_fixed_channel}]
approaches, we present the gauge-invariant reformulation in
Sec.~\ref{subsec:tdcis_vrot}, and a consistent method for evaluating the time derivative of one-electron
observables in Sec.~\ref{subsec:observable}. Then in Sec.~\ref{sec:numerical}
we apply the channel orbital-based TDCIS method, using LG with
fixed orbitals, VG with fixed orbitals, and the reformulated VG, to the
model one-dimensional (1D) Hamiltonian to compare the results of various TDCIS approaches with numerically exact
TDSE results, and demonstrate the importance of non-Ehrenfest method to
compute dipole acceleration. Finally, concluding remarks are given in
Sec.~\ref{sec:conclusion}. The Hartree atomic units are used throughout unless otherwise noted.

\section{Theory\label{sec:theory}}
\subsection{System Hamiltonian and gauge transformation\label{subsec:system}}
Let us consider an atom or a molecule consisting of $N$ electrons interacting
with an external laser field. In this work, we restrict our
treatment in the clamped-nuclei approximation and the electron-laser
interaction within the electric dipole approximation. 
Then the exact description of the system dynamics is
given by the solution $\Psi_{\rm L}(t)$ of TDSE, 
\begin{eqnarray}\label{eq:tdse_1q_lg}
i\dot{\Psi}_{\rm L}(t) = H_{\rm L}(t)\Psi_{\rm L}(t),
\end{eqnarray}
with the system Hamiltonian
$H_{\rm L}(t) = H_0 + H^{\rm ext}_{\rm L}(t)$,
where $H_0$ is the field-free electronic Hamiltonian
\begin{eqnarray}\label{eq:ham0}
H_0 = \sum_{k=1}^N h(\bm{r}_k,\bm{p}_k) + \sum_{k=1}^N\sum_{l>k}^N \frac{1}{|\bm{r}_k-\bm{r}_l|},
\end{eqnarray}
where $\bm{r}_k$ and $\bm{p}_k=-i\bm{\nabla}_k$ are the coordinate and canonical
momentum of an electron,
$h(\bm{r},\bm{p}) = \frac{1}{2}\bm{p}^2+v_{\rm n}(\bm{r})$, with
$v_{\rm n}$ being the electron-nucleus interaction. Here we are
considering the {\color{black}LG} treatment, where the electron-laser interaction 
$H^{\rm ext}_{\rm L}$ is given by
\begin{eqnarray}\label{eq:hamext}
H^{\rm ext}_{\rm L}(t) = \bm{E}(t)\cdot\sum_{k=1}^N \bm{r}_k,
\end{eqnarray}
where $\bm{E}(t)$ is the laser electric field.

As well known, the system dynamics is {\it equivalently} described in
the {\color{black}VG}, of which the wavefunction $\Psi_{\rm V}$ is
connected with the LG one through
\begin{eqnarray}\label{eq:transformation_1q}
\Psi_{\rm V}(t) = U(t)\Psi_{\rm L}(t),
\end{eqnarray}
with a unitary transformation
\begin{eqnarray}\label{eq:uop_1q}
U(t) = \exp\!\left[-i\sum_{k=1}^N\!\left\{\bm{A}(t)\cdot\bm{r}_k-\frac{1}{2}\!\int_{-\infty}^t\!dt^\prime |\bm{A}(t^\prime)|^2\right\}\right], 
\end{eqnarray}
{\color{black} where $\bm{A}(t)=-\int_{-\infty}^t \bm{E}(t^\prime) dt^\prime$ is the vector
potential,} and we arbitrarily include the second term in the exponential, which
is a $c$-number, to avoid appearance of terms proportional to $|\bm{A}|^2$ in subsequent equations. 
Then we substitute $\Psi_{\rm L}=U^{-1}\Psi_{\rm V}$ into the LG TDSE, Eq.~(\ref{eq:tdse_1q_lg}), 
use $dU/dt=i\sum_{k=1}^N(\bm{E}\cdot\bm{r}_k+|\bm{A}|^2/2)U$, and note
$U\bm{p}_kU^{-1}=\bm{p}_k+\bm{A}$ to derive the VG TDSE, 
\begin{eqnarray}\label{eq:tdse_1q_vg}
i\dot{\Psi}_{\rm V}(t) = H_{\rm V}(t)\Psi_{\rm V}(t),
\end{eqnarray}
with $H_{\rm V}(t) = H_0 + H^{\rm ext}_{\rm V}(t)$, and
\begin{eqnarray}\label{eq:hamext}
H^{\rm ext}_{\rm V}(t) = \bm{A}(t)\cdot\sum_{k=1}^N \bm{p}_k.
\end{eqnarray}
One should carefully note that the present proof of equivalence of the LG and VG
treatments, Eqs.~(\ref{eq:tdse_1q_lg}) and (\ref{eq:tdse_1q_vg}), with
the transformation of Eq.~(\ref{eq:uop_1q}), applies only to the exact
solution of TDSE. See e.g, Ref.~\cite{Bandrauk:2009,Han:2010,Bandrauk:2013} for deeper
discussions on the gauge transformation within TDSE, and
Ref.~\cite{Ishikawa:2015} for the gauge invariance of {\color{black}TD-MCSCF} methods.

For a compact presentation of the many-electron theory, we rewrite the 
system Hamiltonian in the second quantization, 
\begin{subequations}\label{eqs:ham2q}
\begin{eqnarray}
\hat{H}_{\rm L}(t)\label{eq:ham2q_lg}
 &=& \hat{H}_0+\hat{H}^{\rm ext}_{\rm L}(t), \\
\hat{H}_{\rm V}(t)\label{eq:ham2q_vg}
 &=& \hat{H}_0+\hat{H}^{\rm ext}_{\rm V}(t),
\end{eqnarray}
\end{subequations}
\begin{eqnarray}\label{eq:H0_2q}
\hat{H}_0 = \hat{h} +
\frac{1}{2}\sum^{\uparrow\downarrow}_{\sigma\tau}\sum_{pqrs}\langle
pr|qs\rangle\hat{c}^\dagger_{p\sigma}\hat{c}^\dagger_{r\tau}\hat{c}_{q\tau}\hat{c}_{s\sigma},
\end{eqnarray}
\begin{subequations}\label{eqs:vext_2q}
\begin{eqnarray}
\hat{H}^{\rm ext}_{\rm L}(t) &=&\label{eq:vext_2q_lg}
\bm{E}(t)\cdot\hat{r}, \\
\hat{H}^{\rm ext}_{\rm V}(t) &=&\label{eq:vext_2q_vg}
\bm{A}(t)\cdot\hat{p},
\end{eqnarray}
\end{subequations}
where $\{\hat{c}^\dagger_{p\sigma}\}$ and $\{\hat{c}_{p\sigma}\}$ are the creation 
and annihilation operators, respectively, for the set of spin-orbitals given as
a direct product $\{\phi_p\}\otimes\{s_\uparrow,s_\downarrow\}$ of
orthonormal spatial orbitals $\{\phi_p\}$ and up-spin (down-spin) 
functions $s_\uparrow$ ($s_\downarrow$). The operators $\hat{h}$, $\hat{\bm{r}}$, and $\hat{\bm{p}}$ are defined, respectively, as
$\hat{h} = \sum_{\sigma}^{\uparrow\downarrow}\sum_{pq}h_{pq}\hat{c}^\dagger_{p\sigma}\hat{c}_{q\sigma}$,
$\hat{\bm{r}}=\sum_\sigma^{\uparrow\downarrow}\sum_{pq}\bm{r}_{pq}\hat{c}^\dagger_{p\sigma}\hat{c}_{q\sigma}$, and
$\hat{\bm{p}}=\sum_\sigma^{\uparrow\downarrow}\sum_{pq}\bm{p}_{pq}\hat{c}^\dagger_{p\sigma}\hat{c}_{q\sigma}$, where
$h_{pq}$, $\bm{r}_{pq}$, and
$\bm{p}_{pq}$ are the matrix elements of $h$, $\bm{r}$,
$\bm{p}$, respectively, in terms of $\{\phi_p\}$, and
\begin{eqnarray}\label{eq:eris}
\langle pr|qs\rangle = \int \!\!d\bm{r}_1\!\!\int \!\!d\bm{r}_2 \phi^*_p({\color{black}\bm{r}_1})\phi^*_r({\color{black}\bm{r}_2})r^{-1}_{12}\phi_q({\color{black}\bm{r}_1})\phi_s({\color{black}\bm{r}_2}). \nonumber \\
\end{eqnarray}
The TDSE of the LG, Eq.~(\ref{eq:tdse_1q_lg}), and VG,
Eq.~(\ref{eq:tdse_1q_vg}), read
\begin{subequations}\label{eqs:tdse_2q} 
\begin{eqnarray}
i|\dot{\Psi}_{\rm L}(t)\rangle &=&\label{eq:tdse_2q_lg} \hat{H}_{\rm L}(t)|\Psi_{\rm L}(t)\rangle, \\
i|\dot{\Psi}_{\rm V}(t)\rangle &=&\label{eq:tdse_2q_vg} \hat{H}_{\rm V}(t)|\Psi_{\rm V}(t)\rangle,
\end{eqnarray}
\end{subequations}
with the transformation 
\begin{eqnarray}\label{eq:transformation_2q}
|\Psi_{\rm V}\rangle=\hat{U}(t)|\Psi_{\rm L}\rangle,  
\end{eqnarray}
\begin{eqnarray}\label{eq:uop_2q}
\hat{U}(t) = \exp\!\left[-i\!\left\{\bm{A}(t)\cdot\hat{\bm{r}}-\frac{\color{black}\hat{N}}{2}\!\int^t_{\color{black}-\infty}\!dt^\prime |\bm{A}(t^\prime)|^2\right\}\right],
\end{eqnarray}
{\color{black}where $\hat{N}=\sum_{\mu}\sum_{\sigma}^{\uparrow\downarrow}\hat{c}^\dagger_{\mu\sigma}\hat{c}_{\mu\sigma}$ is the
number operator.}

In this work, {\color{black}we consider a closed-shell system with even number of electrons, and}
choose as $\{\phi_p\}$ the time-independent Hartree-Fock (HF) orbitals satisfying the
canonical, {\color{black}restricted} HF equation
\begin{eqnarray}
\hat{f} |\phi_p\rangle &\equiv&
\hat{h} |\phi_p\rangle + 2\sum_j\hat{W}^{\phi_j}_{\phi_j} |\phi_p\rangle - \sum_j\hat{W}^{\phi_j}_{\phi_p} |\phi_j\rangle \nonumber \\
&=& \epsilon_p |\phi_p\rangle,
\end{eqnarray}
where $\epsilon_p$ is the orbital energy, and $\hat{W}^\phi_{\phi^\prime}$ is the
electrostatic potential of a product $\phi^{*}(\bm{r})\phi^{\prime}(\bm{r})$ of given orbitals, defined in the real space as
\begin{eqnarray}
W^{\phi}_{\phi^{\prime}}(\bm{r}_1) = \int d\bm{r}_2  \frac{\phi^{*}(\bm{r}_2)\phi^\prime(\bm{r}_2)}{|\bm{r}_1-\bm{r}_2|}.
\end{eqnarray}
As usual, we separate the full set of HF orbitals $\{\phi_p\}$ into
the occupied orbitals $\{\phi_i\}$ which are occupied in the HF ground-state
wavefunction (also referred to as the reference) $|\Phi\rangle=\prod_i
\hat{c}^\dagger_{i\uparrow}\hat{c}^\dagger_{i\downarrow}|\rangle$ ($|\rangle$ is the vacuum.), 
and the virtual orbitals $\{\phi_a\}$ which are unoccupied in $|\Phi\rangle$.

\subsection{Review of CI coefficient-based TDCIS with fixed orbitals\label{subsec:tdcis_fixed_cicoeff}}
We write the second-quantized version of Eq.~(\ref{eq:tdcis_1q}), for the LG case, as
\begin{eqnarray}\label{eq:tdcisl}
|\Psi_{\rm L}(t)\rangle &=& |\Phi\rangle C_0(t) + \sum_i^{occ}\sum_a^{vir}  |\Phi_{ia}\rangle C_{ia}(t),
\end{eqnarray}
where
$|\Phi_{ia}\rangle=\sum_\sigma^{\uparrow\downarrow}\hat{c}^\dagger_{a\sigma}\hat{c}_{i\sigma}|\Phi\rangle/\sqrt{2}$.
The equations of motion for the CI coefficients have been derived \cite{Rohringer:2006} by
inserting Eq.~(\ref{eq:tdcisl}) into the LG TDSE, Eq.~(\ref{eq:tdse_2q_lg}), and closing
from the left with the reference and singly-excited CSFs,
\begin{subequations}\label{eqs:deltasl}
\begin{eqnarray}
\langle\Phi|(\hat{H}_{\rm L}\!-\!i\partial_t)\{|\Phi\rangle C_0
\!+\! \sum_{jb} |\Phi_{jb}\rangle C_{jb}\} &=& \label{eqs:tdsecis_0}0, \\
\langle\Phi_{ia}|(\hat{H}_{\rm L}-\!i\partial_t)\{|\Phi\rangle C_0
\!+\! \sum_{jb} |\Phi_{jb}\rangle C_{jb}\} &=&\label{eqs:tdsecis_ia}0.
\end{eqnarray}
\end{subequations}
{\color{black}Conceptually more proper} derivation of Eqs.~(\ref{eqs:deltasl}) is based on Dirac-Frenkel
variational principle, which considers the Lagrangian
\begin{eqnarray}\label{eq:lagl}
L_{\rm L}(t) = \langle\Psi_{\rm L}|(\hat{H}_{\rm L}\!-\!i\partial_t)|\Psi_{\rm L}\rangle,
\end{eqnarray}
and requires $\partial L_{\rm L}/\partial C^*_0 = \partial L_{\rm L}/\partial C^{*}_{ia} = 0$.
Substituting $\hat{H}_{\rm L}$ of Eq.~(\ref{eq:ham2q_lg}) into Eqs.~(\ref{eqs:deltasl}), 
using the Slater-Condon rule for the Hamiltonian matrix elements, and
noting the canonical condition $f_{pq}=\epsilon_p\delta_{pq}$, the
EOMs for the length gauge are derived as \cite{Rohringer:2006}
\begin{subequations}\label{eqs:eom_cicoeff_lg}
\begin{eqnarray}
i\dot{C}_0 \label{eq:tdcis2lg0}&=& \sqrt{2}\bm{E}\cdot\sum_{jb}\langle\phi_j|\hat{\bm{r}}|\phi_b\rangle C_{jb}, \\
i\dot{C}_{ia} &=&  \langle
\phi_a|\{\sum_{b}(\hat{F}_i+\bm{E}\cdot\hat{\bm{r}})|\phi_b\rangle C_{ib}
+ \sqrt{2}\bm{E}\cdot\hat{\bm{r}}|\phi_i\rangle C_0\} \nonumber \\
&-& \label{eq:tdcis2lg1}\bm{E}\sum_j C_{ja} \cdot\langle\phi_j|\hat{\bm{r}}|\phi_i\rangle.
\end{eqnarray}
\end{subequations}
where the action of the operator $\hat{F}_i$ on a given orbital $\phi$ is defined as
\begin{eqnarray}\label{eq:gfock}
 \hat{F}_i|\phi\rangle = (\hat{f}-\epsilon_i)|\phi\rangle + \sum_j (2\hat{W}^{\phi_j}_{\phi}|\phi_i\rangle-\hat{W}^{\phi_j}_{\phi_i}|\phi\rangle).
\end{eqnarray}

Reference{\color{black}s}~\cite{Krebs:2014,Karamatskou:2014,Chen:2015,Tilley:2015,Karamatskou:2015,Goetz:2016,Karamatskou:2017,Karamatskou:2017b} have used the
same expansion in terms of fixed CSFs also in the VG case,
\begin{eqnarray}\label{eq:tdcisv}
|\Psi_{\rm V}(t)\rangle &=& |\Phi\rangle D_0(t) + \sum_i^{occ}\sum_a^{vir}  |\Phi_{ia}\rangle D_{ia}(t),
\end{eqnarray}
and required Eqs.~(\ref{eqs:deltasl}) to hold, with $\hat{H}_{\rm L}$,
$C_0$, and $C_{ia}$ replaced with $\hat{H}_{\rm V}$, $D_0$, and
$D_{ia}$. This is equivalent to consider the following Lagrangian, 
\begin{eqnarray}\label{eq:lagv}
L_{\rm V}(t) = \langle\Psi_{\rm V}|(\hat{H}_{\rm V}\!-\!i\partial_t)|\Psi_{\rm V}\rangle,
\end{eqnarray}
and to require $\partial L_{\rm V}/\partial D^*_0 = \partial L_{\rm
V}/\partial D^{*}_{ia} = 0$, which derives
\begin{subequations}\label{eqs:eom_cicoeff_vg}
\begin{eqnarray}
i\dot{D}_0 \label{eq:tdcis2vg0}&=& \sqrt{2}\bm{A}\cdot\sum_{jb}\langle\phi_j|\hat{\bm{p}}|\phi_b\rangle D_{jb}, \\
i\dot{D}_{ia} &=&  \langle
\phi_a|\{\sum_{b}(\hat{F}_i+\bm{A}\cdot\hat{\bm{p}})|\phi_b\rangle D_{ib}
+ \sqrt{2}\bm{A}\cdot\hat{\bm{p}}|\phi_i\rangle D_0\} \nonumber \\
&-& \label{eq:tdcis2vg1}\bm{A}\sum_j D_{ja} \cdot\langle\phi_j|\hat{\bm{p}}|\phi_i\rangle.
\end{eqnarray}
\end{subequations}

\subsection{Review of Channel orbital-based TDCIS with fixed orbitals\label{subsec:tdcis_fixed_channel}}
An interesting reformulation of the above-described TDCIS method, as
mentioned in Sec.~\ref{sec:introduction}, has been
proposed in Ref.~\citenum{Rohringer:2006}, which introduces the
time-dependent channel orbitals $|\chi_i\rangle$ that collects all the
single excitations originating from an occupied orbital $|\phi_i\rangle$, 
\begin{eqnarray}\label{eq:chi}
|\chi_i\rangle = \sum_a |\phi_a\rangle C_{ia}(t),
\end{eqnarray}
and rewrites the EOMs in terms of $C_0$ and $\{|\chi_i\rangle\}$ as
\begin{subequations}\label{eqs:eom_channel_lg}
\begin{eqnarray}
i\dot{C}_0 \label{eq:eom_channel_lg_c0}&=& \sqrt{2}\bm{E}\cdot\sum_{j}\langle\phi_j|\hat{\bm{r}}|\chi_j\rangle, \\
i|\dot{\chi}_i\rangle &=& \hat{P}
\{(\hat{F}_i+\bm{E}\cdot\hat{\bm{r}})|\chi_i\rangle
+ \sqrt{2}\bm{E}\cdot\hat{\bm{r}}|\phi_i\rangle C_0\} \nonumber \\
&-& \label{eq:eom_channel_lg_chi}\sum_j |\chi_j\rangle\langle\phi_j|\bm{E}\cdot\hat{\bm{r}}|\phi_i\rangle, 
\end{eqnarray}
\end{subequations}
where $\hat{P}=\hat{1}-\sum_j|\phi_j\rangle\langle\phi_j|$. According to
these EOMs and the initial
conditions [$C_0(t\rightarrow -\infty)=1$, and $\{C_{ia}(t\rightarrow -\infty)=0\} \Longleftrightarrow
\{\chi_i(t\rightarrow -\infty) \equiv 0\}$], the channel orbitals
$|\chi_i\rangle$ gets gradually populated along with the laser-electron
interaction, measuring an excitation of an electron out of
$|\phi_i\rangle$. See Ref.~\cite{Rohringer:2006} for 
interesting properties of the channel orbitals. 

It is also possible to formulate the channel orbital-based scheme based
on the velocity gauge TDCIS using fixed orbitals, although not previously
considered. We, therefore, introduce the analogous quantity
\begin{eqnarray}\label{eq:eta}
|\eta_i\rangle = \sum_a |\phi_a\rangle D_{ia}(t),
\end{eqnarray}
and rewrite Eqs.~(\ref{eqs:eom_cicoeff_vg}) as
\begin{subequations}\label{eqs:eom_channel_vg}
\begin{eqnarray}
i\dot{D}_0 \label{eq:eom_channel_vg_c0}&=& \sqrt{2}\bm{A}\cdot\sum_{j}\langle\phi_j|\hat{\bm{p}}|\eta_j\rangle, \\
i|\dot{\eta}_i\rangle &=& \hat{P}
\{(\hat{F}_i+\bm{A}\cdot\hat{\bm{p}})|\eta_i\rangle
+ \sqrt{2}\bm{A}\cdot\hat{\bm{p}}|\phi_i\rangle D_0\} \nonumber \\
&-& \label{eq:eom_channel_vg_chi}\sum_j |\eta_j\rangle\langle\phi_j|\bm{A}\cdot\hat{\bm{p}}|\phi_i\rangle.
\end{eqnarray}
\end{subequations}
Hereafter, we refer to the method based on
Eqs.~(\ref{eqs:eom_channel_lg}), i.e, the channel orbital-based TDCIS
in the length gauge with fixed orbitals, simply as LG method, and that
based on Eqs.~(\ref{eqs:eom_channel_vg}), i.e, the channel
orbital-based TDCIS in the velocity gauge with fixed orbitals, as VG
method, for notational brevity. 

\subsection{Channel orbital-based TDCIS in the velocity gauge with rotated orbitals\label{subsec:tdcis_vrot}}
The gauge dependence of the LG and VG treatments,
Eqs.~(\ref{eqs:eom_channel_lg}) and (\ref{eqs:eom_channel_vg}),
results from the fact that the ansatz of Eqs.~(\ref{eq:tdcisl}) and
(\ref{eq:tdcisv}), both using fixed orbitals, cannot be connected with
the transformation, Eq.~(\ref{eq:uop_2q}), as is generally the case
for truncated CI expansion using fixed orbitals. For a
method to be gauge invariant, the underlying Lagrangian in LG and VG
cases should be numerically the same when evaluated with the solution of
respective EOMs, which does not hold in the present case, $L_{\rm L}(t)
\neq L_{\rm V}(t)$, with Eqs.~(\ref{eq:lagl}) and (\ref{eq:lagv}). 

Thus we {\it define} the total wavefunction $|\Psi^\prime_{\rm
V}(t)\rangle$, {\color{black}transformed from $|\Psi_{\rm L}(t)\rangle$ to
the velocity gauge},
as
\begin{eqnarray}\label{eq:tdcisvnew}
|\Psi^\prime_{\rm V}(t)\rangle &=& \hat{U}(t)|\Psi_{\rm L}(t)\rangle \nonumber \\ &=&
|\Phi^\prime\rangle C_0(t) + \sum_i^{occ}\sum_a^{vir} |\Phi^\prime_{ia}\rangle C_{ia}(t),
\end{eqnarray}
with $|\Psi_{\rm L}(t)\rangle$ constructed with the solution of
CI coefficient-based EOMs in the LG, Eqs.~(\ref{eqs:eom_cicoeff_lg}). Here $|\Phi^\prime\rangle=\hat{U}(t)|\Phi\rangle$ and
$|\Phi^\prime_{ia}\rangle=\hat{U}(t)|\Phi_{ia}\rangle=\sum_{\sigma}\hat{c}^{\prime\dagger}_{a\sigma}\hat{c}^{\prime}_{i\sigma}|\Phi^\prime\rangle/\sqrt{2}$
are the reference and singly-excited CSF constructed with unitary
rotated orbitals, i.e, $|\phi^\prime_p\rangle=\hat{U}|\phi_p\rangle$ and
$\hat{c}^{\prime}_{p\sigma}=\hat{U}(t)\hat{c}_{p\sigma}\hat{U}^{-1}(t)$.
{\color{black}It should be noted that $|\Psi^\prime_{\rm V}\rangle$ cannot
be rewritten into the form of Eq.~(\ref{eq:tdcisv}) in general.}
{\color{black}Associated with} this wavefunction, we consider the following Lagrangian, 
\begin{eqnarray}\label{eq:lagv2}
L^\prime_{\rm V}(t) = \langle\Psi^\prime_{\rm V}|(\hat{H}_{\rm
 V}\!-\!i\partial_t)|\Psi^\prime_{\rm V}\rangle. 
\end{eqnarray}
The equivalence of this approach to the LG treatment is readily
confirmed by seeing 
\begin{eqnarray}\label{eq:lagv2_equivalence}
L^\prime_{\rm V}(t) &=&
\langle\Psi_{\rm L}|\hat{U}^{-1}(\hat{H}_{\rm V}\!-\!i\partial_t)\hat{U}|\Psi_{\rm L}\rangle \nonumber \\
&=& \langle\Psi_{\rm L}|(\hat{H}_{\rm L}\!-\!i\partial_t)|\Psi_{\rm L}\rangle = L_{\rm L}(t).
\end{eqnarray}

One may naively expect that $L^\prime_{\rm V}$ of Eq.~(\ref{eq:lagv2}),
which differs from $L_{\rm V}$ of Eq.~(\ref{eq:lagv}) only by the replacement of $\Psi_{\rm V}$
with $\Psi^\prime_{\rm V}$, leads to the EOMs of
Eqs.~(\ref{eqs:eom_cicoeff_vg}) with ${\color{black}D_0,\{D_{ia}\},}\{\phi_p\}$ replaced with
${\color{black}C_0,\{C_{ia}\},}\{\phi^\prime_p\}$. This is not the case, however, due to the time
dependence of the rotated CSFs, e.g,
$\langle\Phi^\prime|\dot{\Phi}^\prime_{ia}\rangle =
i\bm{E}(t)\cdot\langle\Phi^\prime|\hat{\bm{r}}|\Phi^\prime_{ia}\rangle$,
and after extracting these time dependence, Eq.~(\ref{eq:lagv2}) reads
\begin{eqnarray}\label{eq:ll_transformed3}
L^\prime_{\rm V}(t) = \langle\Psi^\prime_{\rm V}|\{\hat{H}_{\rm V}+\bm{E}(t)\cdot\hat{\bm{r}}-i\partial^{\rm c}_t\}|\Psi^\prime_{\rm V}\rangle,
\end{eqnarray}
where $\partial^{\rm c}_t$ time differentiates CI coefficients only. 
Now requiring $\partial L^\prime_{\rm V}/\partial C^*_0 = \partial
L^\prime_{\rm V}/\partial C^{*}_{ia} = 0$, or equivalently, substituting
the back transformation
$|\phi_p\rangle=\hat{U}^{-1}|\phi^\prime_p\rangle$ into
Eqs.~(\ref{eqs:eom_cicoeff_lg}) derives
\begin{subequations}\label{eqs:eom_cicoeff_vg2}
\begin{eqnarray}
i\dot{C}_0 \label{eq:tdcis2lg0}&=& \sqrt{2}\bm{E}\cdot\sum_{jb}\langle\phi^\prime_j|\hat{\bm{r}}|\phi^\prime_b\rangle C_{jb}, \\
i\dot{C}_{ia} &=&  \langle
\phi^\prime_a|\{\sum_{b}(\hat{F}^\prime_i+\bm{A}\cdot\hat{\bm{p}}+\bm{E}\cdot\hat{\bm{r}})|\phi^\prime_b\rangle C_{ib} \nonumber \\
&+& \sqrt{2}\bm{E}\cdot\hat{\bm{r}}|\phi^\prime_i\rangle C_0\} \nonumber \\
&-& \label{eq:tdcis2lg1}\bm{E}\sum_j C_{ja} \cdot\langle\phi^\prime_j|\hat{\bm{r}}|\phi^\prime_i\rangle.
\end{eqnarray}
\end{subequations}
where $\hat{F}^\prime_i$ is given by Eq.~(\ref{eq:gfock}) with $\{\phi_j\}$ replaced
with $\{\phi^\prime_j\}$. Equations~(\ref{eqs:eom_cicoeff_vg2}) are the CI
coefficient-based TDCIS EOMs based on the Lagrangian of
Eq.~(\ref{eq:lagv2}). Although this approach is guaranteed to be
equivalent to the CI coefficient-based LG TDCIS,
it brings no numerical gain over Eqs.~(\ref{eqs:eom_cicoeff_lg}),
peculiarly including both $\bm{E}\cdot\bm{r}$ and $\bm{A}\cdot\bm{p}$,
and requiring extensive gauge transformation of all occupied and virtual
orbitals. 

None the less, a useful method can be derived, if one switches to the channel orbital-based scheme by
defining the rotated channel functions,
\begin{eqnarray}\label{eq:channelrot}
|\chi^\prime_i(t)\rangle = \hat{U}(t) |\chi_i\rangle =
 \sum_a|\phi^\prime_a\rangle C_{ia}.
\end{eqnarray}
Then we use $d\hat{U}/dt=i(\bm{E}\cdot\bm{\hat{r}}+{\color{black}\hat{N}}|\bm{A}|^2/2)\hat{U}$, and note $\hat{U}\bm{\hat{p}}\hat{U}^{-1}=\bm{\hat{p}}+{\color{black}\hat{N}}\bm{A}$ to derive
\begin{subequations}\label{eqs:eom_channel_vg2}
\begin{eqnarray}
i\dot{C}_0 \label{eq:eom_channel_vg2_c0}&=& \sqrt{2}\bm{E}\cdot\sum_{j}\langle\phi^\prime_j|\hat{\bm{r}}|\chi^\prime_j\rangle, \\
i|\dot{\chi}^\prime_i\rangle &=& \hat{P}^\prime
\{(\hat{F}^\prime_i+\bm{A}\cdot\hat{\bm{p}})|\chi^\prime_i\rangle
+ \sqrt{2}\bm{E}\cdot\hat{\bm{r}}|\phi^\prime_i\rangle C_0\} \\
&-& \label{eq:eom_channel_vg2_chi}\sum_j (
|\chi^\prime_j\rangle\langle\phi^\prime_j|\bm{E}\cdot\hat{\bm{r}}|\phi^\prime_i\rangle+
|\phi^\prime_j\rangle\langle\phi^\prime_j|\bm{A}\cdot\hat{\bm{p}}|\chi^\prime_i\rangle
), \nonumber 
\end{eqnarray}
\end{subequations}
where $\hat{P}^\prime=1-\sum_j|\phi^\prime_j\rangle\langle\phi^\prime_j|$.
Equations~(\ref{eqs:eom_channel_vg2}) are the main
results of this work, which are called the rotated velocity-gauge (rVG) EOMs for brevity.
The rVG scheme is equivalent to the LG scheme with fixed orbitals by construction, while replacing the
{\color{black}length-gauge} dipole operator $\bm{E}\cdot\hat{\bm{r}}$ [the second term of
Eq.~(\ref{eq:eom_channel_lg_chi})] with the spatially uniform
$\bm{A}\cdot\hat{\bm{p}}$ [the second term of Eq.~(\ref{eq:eom_channel_vg2_chi})].
Although several terms in the EOMs still involve the dipole operator, they all apply to the (rotated) occupied orbital which is localized around nuclei,
thus posing no difficulty in enjoying the same advantages of VG propagations of orbitals \cite{Nurhuda:1999,Grum-Grzhimailo:2010,Sato:2016}.

\subsection{Evaluation of the time derivative of an observable\label{subsec:observable}}
Let us next consider how to compute expectation value of a one-electron
operator
$\langle\hat{O}\rangle(t)=\langle\Psi(t)|\hat{O}|\Psi(t)\rangle$, and
its time derivative $d\langle\hat{O}\rangle/dt$. 
For exact solution of TDSE, $\dot{|\Psi\rangle}=-i\hat{H}|\Psi\rangle$,
the time derivative is given by
\begin{subequations}\label{eqs:tdop1e}
\begin{eqnarray}
\frac{d}{dt}\langle\Psi|\hat{O}|\Psi\rangle &=&
 \langle\Psi|\hat{O}|\dot{\Psi}\rangle +
 \langle\dot{\Psi}|\hat{O}|\Psi\rangle \label{eq:tdop1e_td} \\
&=& -i \langle\Psi|[\hat{O},\hat{H}]|\Psi\rangle, \label{eq:tdop1e_ehrenfest}
\end{eqnarray}
\end{subequations}
known as the Ehrenfest expression. For an approximate method, however, the Ehrenfest theorem,
Eq.~(\ref{eq:tdop1e_ehrenfest}), {\color{black}generally does not} hold, and one should
explicitly evaluate the time derivative as
Eq.~(\ref{eq:tdop1e_td}). Important exceptions include those theories using time-dependent
orbitals evolving to satisfy the time-dependent variational principle,
such as {\color{black}time-dependent Hartree-Fock (TDHF)}, TD-MCSCF,
and {\color{black}time-dependent density functional theory}. See Ref.~\cite{Sato:2016} for more details. 
\begin{widetext} 
The TDCIS expectation value of a one-electron operator $\hat{O}$ is
given \cite{Rohringer:2006} by
\begin{eqnarray}\label{eq:op1e}
&&\langle\Psi_{\rm L}|\hat{O}|\Psi_{\rm L}\rangle = 2\sum_j\langle\phi_j|\hat{O}|\phi_j\rangle +\sum_j\langle\chi_j|\hat{O}|\chi_j\rangle
+2\sqrt{2}\operatorname{Re}{[C^*_0\sum_j\langle\phi_j|\hat{O}|\chi_j\rangle]}
- \sum_{ij}\langle\chi_i|\chi_j\rangle
\langle\phi_j|\hat{O}|\phi_i\rangle, \nonumber \\
\end{eqnarray}
in the LG case. That for the VG is given by replacing $C_0$ with $D_0$
in the above equation, and for the rVG by replacing $\{\phi_j,\chi_j\}$ with $\{\phi^\prime_j,\chi^\prime_j\}$.
The expression for the time derivative, in the LG case, is derived by
using Eqs.~(\ref{eqs:eom_channel_lg}) in Eq.~(\ref{eq:tdop1e_td}) as
\begin{eqnarray}\label{eq:dtop1efixed}
\frac{d \langle\Psi_{\rm L}|\hat{O}|\Psi_{\rm L}\rangle}{dt} = 2\operatorname{Re}\left[
\sum_j \langle\chi_j|\hat{O}|\dot{\chi}_j\rangle
+ \sqrt{2}(\dot{C}^*_0\langle\phi_j|\hat{O}|\chi_j\rangle +
C^*_0\langle\phi_j|\hat{O}|\dot{\chi}_j\rangle)
-\sum_{ij}\langle\chi_i|\dot{\chi}_j\rangle\langle\phi_j|\hat{O}|\phi_i\rangle
\right].
\end{eqnarray}
The VG expression is also given by the above equation with $C_0$ replaced
with $D_0$, and that for the rVG is
\begin{eqnarray}\label{eq:dtop1erot}
&&\frac{d \langle\Psi^\prime_{\rm V}|\hat{O}|\Psi^\prime_{\rm V}\rangle}{dt} = 2\operatorname{Re}\left[
\sum_j \langle\chi^\prime_j|\hat{O}|\dot{\chi}^\prime_j\rangle
+\sqrt{2}(\dot{C}^*_0\langle\phi^\prime_j|\hat{O}|\chi^\prime_j\rangle +
C^*_0\langle\phi^\prime_j|\hat{O}|\dot{\chi}^\prime_j\rangle)
-\sum_{ij}\langle\chi^\prime_i|\dot{\chi}^\prime_j\rangle\langle\phi^\prime_j|\hat{O}|\phi^\prime_i\rangle
\right] \\
&&\quad+ \sqrt{2}\operatorname{Im}\left[
2\bm{E}\cdot\sum_jC^*_0\langle\phi^\prime_j|\hat{\bm{r}}\hat{O}|\chi^\prime_j\rangle
+|\bm{A}|^2 \sum_j C^*_0\langle\phi^\prime_j|\hat{O}|\chi^\prime_j\rangle
\right]
- i\bm{E}\cdot\sum_{ij}(2\delta_{ij}-\langle\chi^\prime_i|\chi^\prime_j\rangle)
\langle\phi^\prime_j|[\hat{\bm{r}},\hat{O}]|\phi^\prime_i\rangle. \nonumber
\end{eqnarray}
Although Eqs~(\ref{eq:dtop1efixed}) and (\ref{eq:dtop1erot}) look rather
complicated, their evaluations are straightforward given the time
derivatives of working variables $C_0$, $\{\chi_i\}$, etc, which are
necessary, in any case, to propagate the EOMs.
\end{widetext}

\begin{figure}[!t]
\centering
\includegraphics[width=0.9\linewidth,clip]{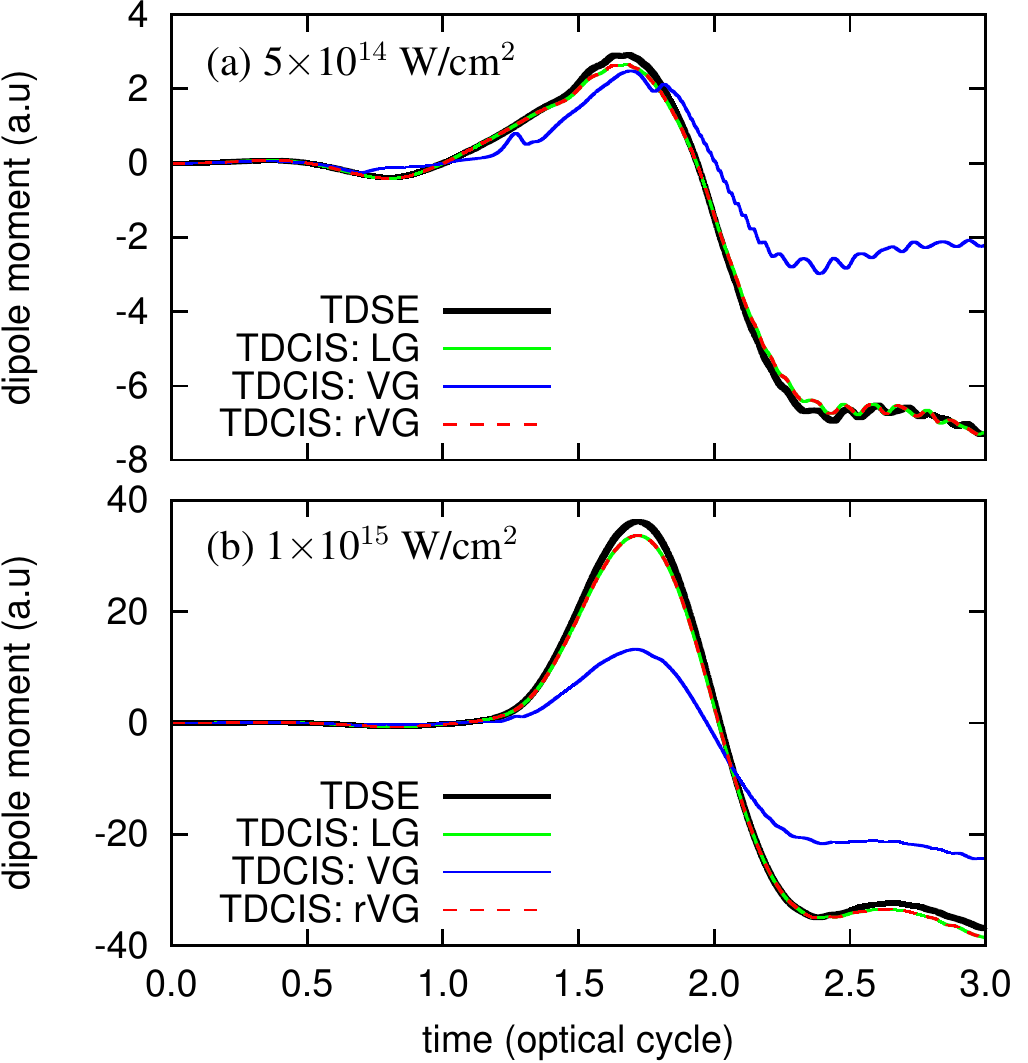}
\caption{\label{fig:dip}Time evolution of the dipole moment of 1D-He
exposed to a laser pulse with a wavelength of 750 nm and an intensity
of (a) 5$\times$10$^{14}$ W/cm$^2$ and (b) 1$\times$10$^{15}$ W/cm$^2$.
Comparison of the results with TDCIS in the LG, VG, and rVG
 with that of TDSE.}%
\end{figure}


\section{Numerical examples}\label{sec:numerical}
In this section, we numerically apply the channel orbital-based TDCIS method in the LG, VG, and rVG to the 1D model Helium atom, 
using the computational code developed by modifying an existing TDHF code used in our previous work \cite{Sato:2013,Sato:2014,Sato:2015}.
The field-free electronic Hamiltonian is given by
\begin{eqnarray}
H_0 = \sum_{k=1}^2 \left\{-\frac{1}{2}\frac{\partial^2}{\partial z^2_k}-\frac{2}{z^2_k+1}\right\} + \frac{1}{\sqrt{(z_1-z_2)^2+1}}, \nonumber \\
\end{eqnarray}
for two electronic coordinates $z_1$ and $z_2$, and the laser-electron interaction
$\bm{E}(t)\cdot\bm{r}$ and $\bm{A}(t)\cdot\bm{p}$ are replaced with
$E(t)z$ and $A(t)p_z=-iA(t)\partial/\partial_z$, 
respectively, in Eqs.~(\ref{eqs:eom_channel_lg}), 
(\ref{eqs:eom_channel_vg}) and (\ref{eqs:eom_channel_vg2}). 
Orbitals are discretized on equidistant
grid points with spacing $\Delta z=0.4$ within a simulation box
$-1000\leq z\leq 1000$, with an absorbing boundary implemented
by a mask function of $\cos^{1/4}$ shape at 10\% side edges of the
box. Each EOM is solved by the fourth-order Runge-Kutta
method with a fixed time step size (1/10000 of an optical
cycle). Spatial derivatives are evaluated by the eighth order
finite difference method, and spatial integrations are
performed by the trapezoidal rule. 
We consider a laser electric field given by
\begin{eqnarray}\label{eq:laser}
E(t) = E_0 \sin(\omega_0 t) \sin^2\left(\pi \frac{t}{\tau}\right),
\end{eqnarray}
for $0 \leq t \leq \tau$, and $E(t)=0$ otherwise, 
with a wavelength $\lambda=2\pi/\omega_0=750$ nm, a foot-to-foot pulse length $\tau$ of three optical cycles,
and a peak intensity $I_0=E_0^2$ for $I_0=5\times10^{14}$ W/cm\ue{2} and $I_0=10^{15}$ W/cm\ue{2}.
The 1D Hamiltonian, computational details, and the applied laser field are the same as used in 
Ref.~\cite{Sato:2014} to facilitate comparison with TDSE results in Ref.~\cite{Sato:2014}.

\begin{figure}[!t]
\centering
\includegraphics[width=0.9\linewidth,clip]{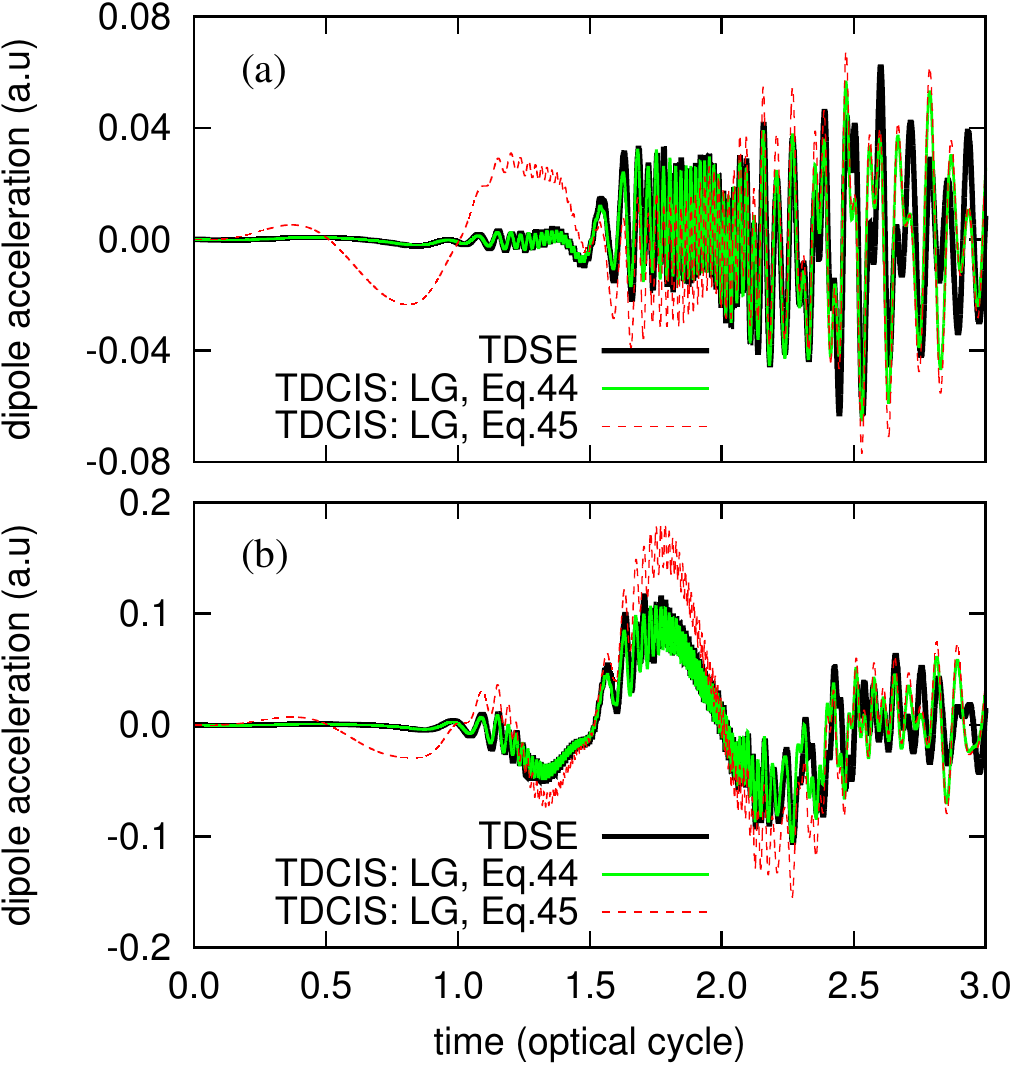}
\caption{\label{fig:acc_ehrenfest}Time evolution of the dipole acceleration of 1D-He
exposed to a laser pulse with a wavelength of 750 nm and an intensity
of (a) 5$\times$10$^{14}$ W/cm$^2$ and (b) 1$\times$10$^{15}$ W/cm$^2$.
Comparison of the results with TDCIS in the LG adopting Eq.~(\ref{eq:a_td}) and Eq.~(\ref{eq:a_ehrenfest}) with that of TDSE.}%
\end{figure}
First, we compare the time-dependent dipole moment $\langle z\rangle(t)$
obtained with TDCIS approaches with that of TDSE in Fig.~\ref{fig:dip},
which immediately reveals a strong gauge dependence of fixed-orbital
approaches, i.e, the large {\color{black}difference between} LG and VG results.
One should note that the comparison of LG and VG results alone can tell
nothing about the preference of either approach; TDCIS method in both LG and VG
are the first approximation in the hierarchy
of CI expansions, which, at the full-CI limit, would be gauge
invariant. The point here is that the LG scheme
outperforms the VG scheme in comparison to the exact TDSE result
as clearly seen in Fig.~\ref{fig:dip}, which convinces one an {\it
empirical} preference of the LG treatment. 
On the other hand, the results of LG and rVG agree
perfectly within the graphical resolution, numerically demonstrating the theoretical
gauge invariance.

Next, we consider the dipole acceleration $\langle a\rangle(t)$ defined as the time derivative
of the kinematic momentum, 
\begin{eqnarray}\label{eq:a_td}
\langle a\rangle(t) = \frac{d\langle\hat{\pi}\rangle}{dt}, 
\end{eqnarray}
where $\hat{\pi}=\hat{p}_z$ for the LG, and $\hat{\pi}=\hat{p}_z+A(t)$
for the VG. In the exact
TDSE case, applying Eqs.~(\ref{eqs:tdop1e}) for
$\hat{O}=\hat{\pi}$ (also taking into account the trivial, explicit time dependence of
$\pi(t)$ in the VG case) derives
\begin{eqnarray}\label{eq:a_ehrenfest}
\langle a\rangle(t) =
 -\langle\Psi|\frac{\partial\hat{v}_{\rm nuc}}{\partial\hat{z}}|\Psi\rangle-2E(t),
\end{eqnarray}
where $\partial v_{\rm nuc}/\partial z=-\partial/\partial_z 2(z^2+1)^{-1/2}
= 2z(z^2+1)^{-3/2}$ for the 1D Hamiltonian. Numerically achieving the
theoretical equivalence of Eq.~(\ref{eq:a_td}) and
(\ref{eq:a_ehrenfest}), even for the exact TDSE method, requires a
simulation to be converged with respect to computational parameters (time-step
size, etc). Therefore, we first {\color{black}applied} both
Eq.~(\ref{eq:a_td}) and Eq.~(\ref{eq:a_ehrenfest}) in the TDSE
simulation, and {\color{black}confirmed} a perfect agreement (not shown), suggesting the
convergence of the simulation. Then we compare the results of TDCIS in
the LG, using Eqs.~(\ref{eq:a_td}) [i.e, Eq.~(\ref{eq:dtop1efixed}) with
$\hat{O}=\hat{p}_z$] and (\ref{eq:a_ehrenfest}), with that of TDSE in
Fig.~\ref{fig:acc_ehrenfest},
clearly showing a better agreement of the results of the former approach with
that of TDSE. From this result, and also by the fact that being based on
Eq.~(\ref{eq:a_td}) guarantees that the HHG spectra obtained from the
velocity $\langle\pi\rangle(t)$ and the acceleration $\langle
a\rangle(t)$, at the convergence, properly relate to each other \cite{Bandrauk:2009},
we consider that Eq.~(\ref{eq:a_td}), {\color{black}together with
Eq.~(\ref{eq:dtop1efixed}) or Eq.~(\ref{eq:dtop1erot})}, should be adopted as a {\it
consistent} method for evaluating the dipole acceleration. 
\begin{figure}[!t]
\centering
\includegraphics[width=0.9\linewidth,clip]{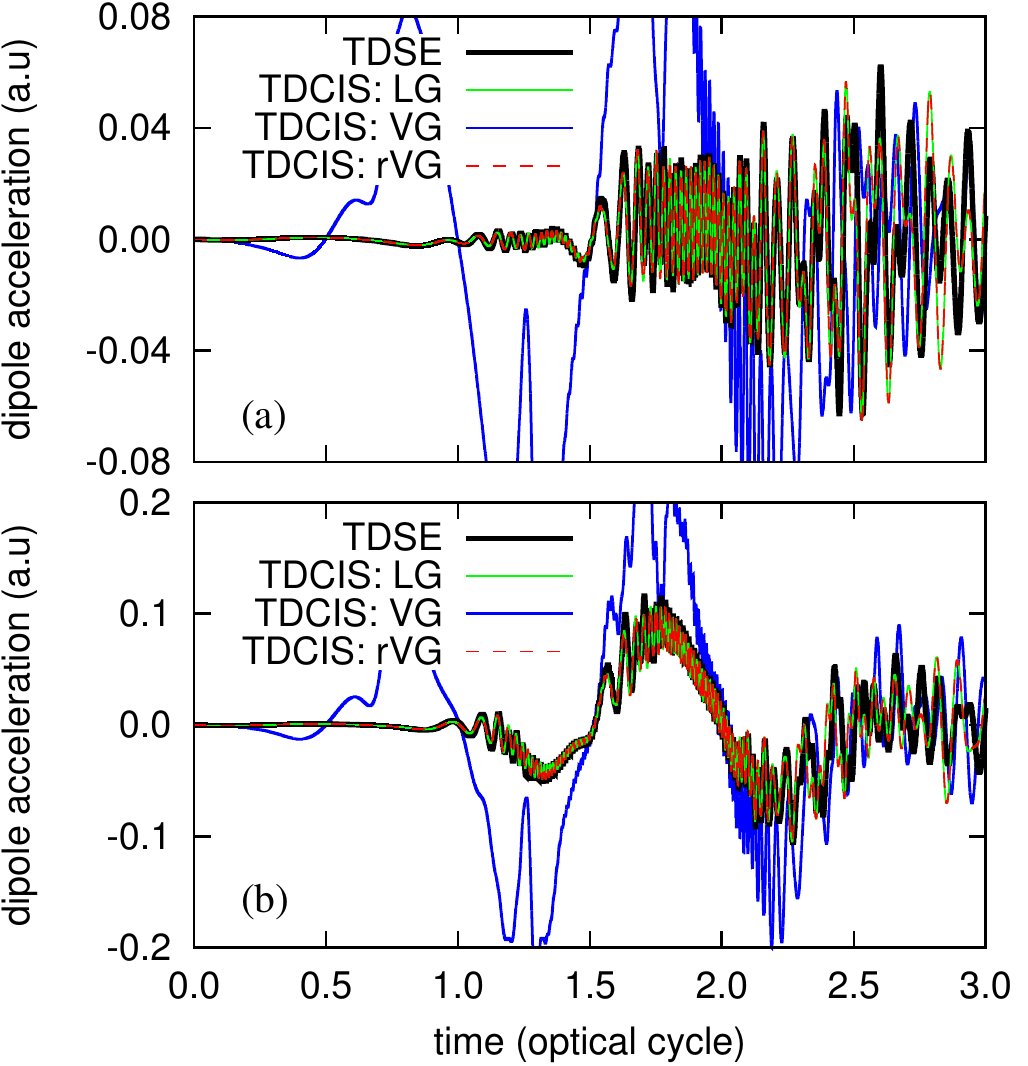}
\caption{\label{fig:acc_gauge}Time evolution of the dipole acceleration of 1D-He
exposed to a laser pulse with a wavelength of 750 nm and an intensity
of (a) 5$\times$10$^{14}$ W/cm$^2$ and (b) 1$\times$10$^{15}$ W/cm$^2$.
Comparison of the results with TDCIS in the LG, VG, and rVG with that of TDSE.}%
\end{figure}

\begin{figure}[!t]
\centering
\includegraphics[width=0.9\linewidth,clip]{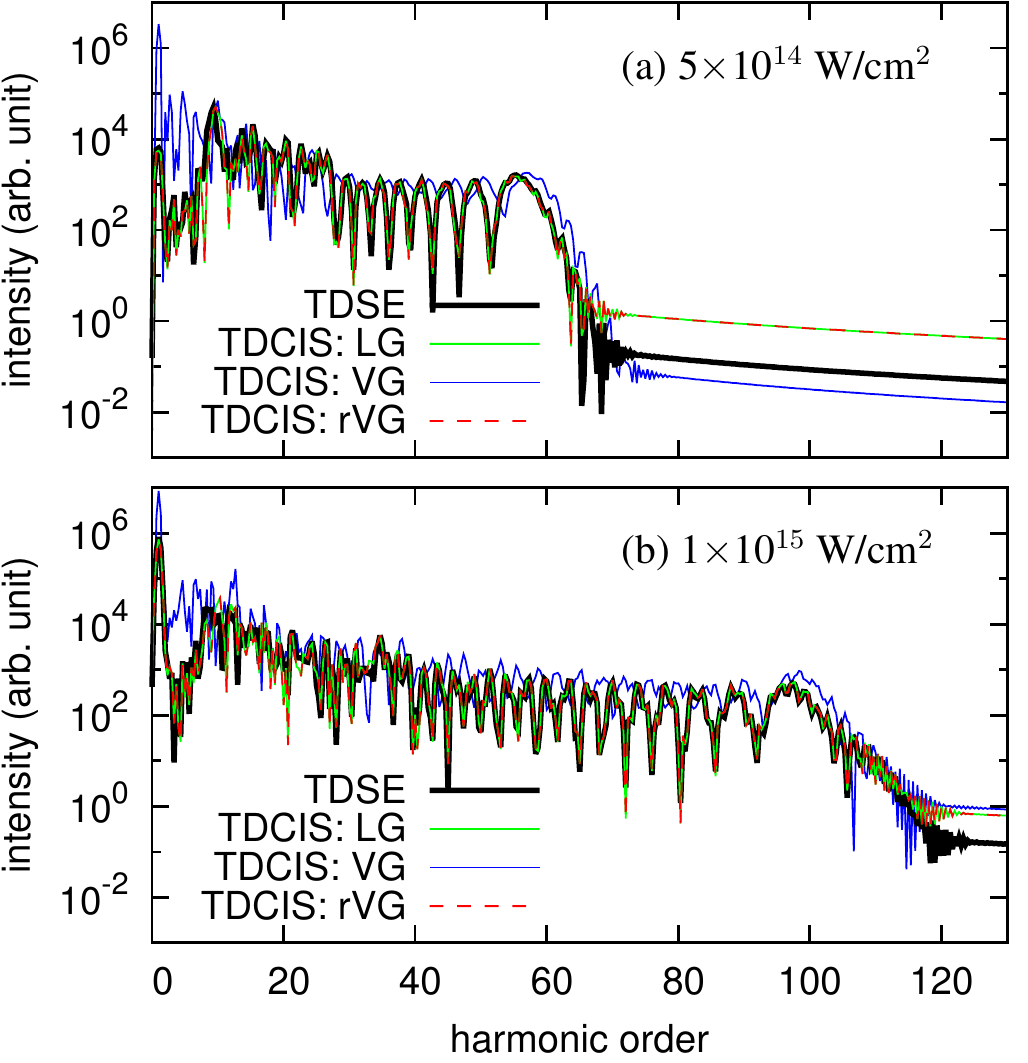}
\caption{\label{fig:hhg}HHG spectrum of 1D-He
exposed to a laser pulse with a wavelength of 750 nm and an intensity
of (a) 5$\times$10$^{14}$ W/cm$^2$ and (b) 1$\times$10$^{15}$ W/cm$^2$.
Comparison of the results with TDCIS in the LG, VG, and rVG
with that of TDSE.}%
\end{figure}
Then we compare the time evolution of the dipole acceleration
[Fig.~\ref{fig:acc_gauge}] and the HHG spectrum [Fig.~\ref{fig:hhg}] obtained as the modulus
squared of the Fourier transform of the dipole acceleration
obtained with TDCIS method in LG, VG, and rVG [based on
Eq.~(\ref{eq:a_td})] with those of TDSE. We observe that (1) the LG and
rVG results are identical to within the 
scale of the figure, (2) they also show a good agreement with TDSE
results, (3) and in contract, the VG results strongly deviate from all
the other results. Especially, Fig.~\ref{fig:hhg} shows a remarkable
agreement of the TDCIS spectra in the LG and rVG and the TDSE one,
suggesting that the TDCIS method would be a useful computational method
for studying HHG process in more complex atoms and molecules, in particular, 
when the present rVG treatment is combined with advanced, velocity
gauge-specific computational techniques.


\section{Conclusions}\label{sec:conclusion}
In this work, we propose a gauge-invariant formulation of the channel
orbital-based TDCIS method for {\it ab initio} investigations of electron dynamics in atoms and molecules.
Instead of using fixed orbitals both in length-gauge and velocity-gauge
simulations, we adopt, in the velocity-gauge case, the EOMs derived with
unitary rotated orbitals $|\phi^\prime_p(t)\rangle=\hat{U}(t)|{\color{black}\phi_p}\rangle$ using gauge-transforming operator $\hat{U}(t)$,
which replaces the length-gauge operator $\bm{E}\cdot\bm{r}$
appearing in the length-gauge EOMs with
the velocity-gauge counterpart $\bm{A}\cdot\bm{p}$, while keeping the
equivalence to the length-gauge treatment. 
This would make it possible to take advantages of the velocity-gauge
simulation over the length-gauge one, e.g, the faster convergence of
simulations of atoms interacting with an intense and/or long-wavelength
laser field, with respect to the maximum angular momentum included to
expand orbitals, and 
the native feasibility of advanced absorbing boundaries such as the
exterior complex scaling. Applications to real atoms and
molecules with the three-dimensional Hamiltonian will be presented elsewhere.

\acknowledgments{
{\color{black}We thank Yuki Orimo for carefully reading the manuscript.}
This research was supported in part by a Grant-in-Aid for Scientific Research 
(Grants No.~26390076, No.~26600111, No.~16H03881, and 17K05070)
from the Ministry of Education, Culture, Sports, Science and Technology (MEXT) of Japan and also 
by the Photon Frontier Network Program of MEXT.
This research was also partially supported by the Center of Innovation Program from the Japan Science 
and Technology Agency, JST, and by CREST (Grant No.~JPMJCR15N1), JST.
Y.~O. gratefully acknowledges support from the Graduate School of
Engineering, The University of Tokyo, Doctoral Student Special
Incentives Program (SEUT Fellowship).
} 

\bibliography{refs.bib}
\end{document}